\documentclass[twocolumn,superscriptaddress,aps,prb,amsmath,preprintnumbers]{revtex4-2}
\usepackage{tikz}
\usepackage{graphicx}
\usepackage{hyperref}
\usepackage{epsfig}
\usepackage{xcolor}
\usepackage{amssymb}

\begin{document}
	
\title{States decoupled from the surface in short Si atomic chains}
		
\author{T.\ Kwapi\'nski}
\email{tomasz.kwapinski@umcs.pl}	
\affiliation{Institute of Physics, M.\ Curie Sk\l odowska 	University, 20-031 Lublin, Poland}		
\author{M. Dachniewicz}
\affiliation{Institute of Physics, M.\ Curie Sk\l odowska 	University, 20-031 Lublin, Poland}	
\author{M. Kurzyna}
\affiliation{Institute of Computer Science and Mathematics, M.\ Curie Sk\l odowska 	University, 20-031 Lublin, Poland}	
\author{M. Ja\l ochowski}
\affiliation{Institute of Physics, M.\ Curie Sk\l odowska 	University, 20-031 Lublin, Poland}	

\date{\today}
	
\begin{abstract}
We analyze both the stationary and time-dependent properties of molecular states in atomic chains on a surface, some of which are composed of atomic states decoupled from the substrate — a phenomenon analogous to dark states in quantum dot systems. To illustrate this effect at the atomic scale, we performed scanning tunneling microscopy (STM) experiments on short silicon chains fabricated on a Si(553)-Au surface.
In contrast to quantum dots, which typically involve characteristic energies in the meV range or lower, the atomic chains studied here operate in a high-energy regime, with energies in the eV range.
Furthermore, we demonstrate that the local density of states of the chains carries clear signatures of these decoupled states, which significantly affect STM imaging. The topography becomes highly sensitive to the bias polarity, to the extent that some atomic sites may appear nearly invisible to the STM tip.
Our time-resolved theoretical analysis reveals that these decoupled states emerge over a finite time interval. This oscillatory dynamical evolution, primarily driven by nearest-neighbor interactions, suggests a universal relaxation mechanism that is largely insensitive to the length of the atomic chain.
\end{abstract}
	
\keywords{decoupled state, molecular state, atomic chain, STM, vicinal surface, Si(553)}
	

\maketitle

\section{\label{sec1}Introduction}

A quantum system consisting of  $N$ atoms possesses at least $N$ molecular eigenstates with distinct energies. In general, these eigenstates are expressed as linear combinations of all atomic states forming the system. However, under certain conditions, a single molecular state may include no contribution from a particular atomic state. As a result, this atomic state becomes decoupled from the environment, rendering the corresponding atom effectively invisible to its surroundings.
The key factor behind this effect is the coherent superposition of atomic states, a phenomenon originally discovered in the fluorescence of sodium atoms \cite{Alzetta1976}, and commonly referred to as dark states. This effect is closely related to various physical phenomena such as coherent adiabatic transitions, electromagnetically induced transparency, and long-distance quantum communication. Systems exhibiting such decoupled states offer a wide range of potential applications, 
for instance in quantum information processing, where they can be used as multi-bit dark-state memory devices, or in advanced cooling techniques. \cite{Aharon2016,Fle2000,An2022,Gry2001}

Recently such states were analyzed also for purely electronic systems such as artificial atoms e.g.  in the stationary triple quantum dot (QD) systems where decoupled states completely blocked the conductance \cite{Michaelis2006, Emary2007, Poltl2009, Kostyrko2009, Weymann2011, Wrzesniewski2018}.  This phenomenon was also studied theoretically in the double QD structure with four separate eigenstates \cite{Pozner2015, Aharon2016}. 
Electron tunneling through such systems is then determined by the interference effects of various electron channels \cite{a01,a02,a07}, which can lead to  the Fano resonance, conductance oscillations, or charge waves. In certain system geometries, these electron paths can cause the effective coupling of a localized state to vanish, resulting in a state that is completely decoupled from the environment (dark states) \cite{a01,a02,Brandes2005}.

Predicting the existence of states decoupled from the leads in QD systems raises the possibility that this phenomenon may also occur in real-atom structures fabricated on a substrate.  
However, in such atomic systems, coupling with the substrate may suppress this effect, preventing it from manifesting.
To the best of our knowledge, states decoupled from the environment in elementary structures composed of  atoms on surfaces have not yet been investigated in the stationary case (i.e., in the absence of external driving fields such as lasers).
This work, therefore, explores the nature of such states in self-assembled short atomic chains fabricated on a stepped substrate. The existence of decoupled states in atomic systems can significantly influence their electrical properties, as these linear chains represent the smallest possible electric conductors.
More importantly, 
in this work we show that these states are responsible for unusual scanning tunneling microscopy (STM) topography images, making some atoms appear invisible in STM experiments . 
Additionally, unlike previous theoretical studies of dark states in QDs, where 
characteristic energies are in the meV range \cite{Volk2013} (or lower), real-atom systems correspond to characteristic energies in the eV range,\cite{Jal2024}  making them far more promising for practical applications.

One of the possible candidates for experimental studies of decoupled-state physics in the atomic scale is the Si chain forming stepped edges in vicinal surfaces of Si(111) stabilized with Au atoms   \cite{JAL1997,Crain2004}. 
This surface with atomic chains was intensively investigated in the literature \cite{Ahn2005,Crain2005,Snijders2006,Ryang2007,Polei2013,Hafke2016,Braun2018,Edler2019,Pfnur2021,Krawiec2010} and it reveals charge waves along Si chains \cite{Yeom2022,Jal2023}, as well as the existence of a ground state with a triple degeneracy that represent an insulator with $\mathbb{Z}_3$ topology with fractional solitons \cite{Yeom2022NN}. 
In this system double metallic Au chain on terraces is mainly responsible for its one-dimensional nature whereas Si chain at step edges have localized or weakly dispersed states \cite{Krawiec2010, Snijders2012, Braun2018, Yeom2022}.    
 In our studies, however, we will concentrate on relatively short chains terminated by vacancy defects along regular chains \cite{Ryang2007}. Such a system is not purely metallic; rather, it is a semiconductor with metallic Au chains on its terraces. Each site in the edge chain can therefore be treated as coupled both to its neighboring sites in the chain and to a separate underlying electrode — analogous to a quantum dot geometry featuring dark states.

In this article, we present a theoretical and experimental investigation aimed at confirming the existence of electronic states decoupled from the substrate in linear atomic chains fabricated on a semiconductor surface. We analyze the electronic properties of these nanostructures with particular emphasis on the local density of states (DOS), STM conductance, and topographic imaging. Our findings demonstrate that these observables provide insight into the molecular states of the atomic system under study, including the identification of states that are effectively decoupled from the underlying substrate.
In this context, it is worth answering the question: \emph{Can decoupled states appear for every atomic chain length on the surface?} And then: \emph{Are they occurring throughout the chain or only at specific atomic places?} Also: \emph{What is the role of these states in topographic images for different STM polarities?} 
In our theoretical and experimental investigation, we tackle these challenges by combining numerical calculations with analytical formulas to predict which systems exhibit decoupled states. Moreover, our STM topography results strongly support these predictions.

Additionally, we aim to investigate the time-dependent phenomena underlying the formation of decoupled states. Instead of assuming their instantaneous emergence, our study focuses on the system’s dynamical evolution, which is expected to exhibit characteristic oscillatory behavior in the local DOS before the decoupled states establish at equilibrium.
The timescales of atomic processes are typically very short, in the range of $10^{-12}-10^{-15}$s, and remain inaccessible to STM-based techniques; however, they are well within reach of advanced optical methods.
Here, by performing a numerical analysis of the evolution operator, we aim to determine the relevant oscillation timescale, thereby addressing the challenge of identifying relaxation mechanisms in such low-dimensional systems.

The paper is organized as follows. In Sec.~\ref{sec2}, the experimental details and topography images for Si chains are shown for two STM polarities. Section \ref{sec3} concerns the theoretical model with the Hamiltonian and the calculation method with analytical solutions and in Sec. \ref{sec3b} experimental and numerical results for short chains with decoupled states are analyzed.  Sec.~\ref{sec4} is devoted to the study of the time dynamics of these states and the last Section gives a short summary.

\section{\label{sec2}Experiment}


The subject of the research was chains of Si atoms on the Si(553)-Au surface. This vicinal surface is composed of Si(111) terraces with a width of 1.45 nm. At each terrace alternately arranged double chains of Au atoms exist that  are mainly responsible for the metallic nature of this surface, whereas 1D Si chains appear at step edges with well localized or weakly dispersed states \cite{Krawiec2010, Snijders2012,Braun2018,Yeom2022} being well suited to the study of decoupled states in atomic chains.
The experiments were carried out in an ultra-high vacuum system with a base pressure in the main chamber in the middle of the range of 10$^{- 11}$ mbar. The system was equipped with Reflection High Energy Electron Diffraction (RHEED), OMICRON VT STM apparatus, gold deposition source and a precise quartz microbalance sensor.  Scanning tunneling measurements were 
conducted over the course of several months on various Si(553)-Au surfaces at temperatures ranging from 77K  to 110 K. 
N-type Si (553) samples with a specific resistivity of 0.002 $ \div $ 0.01 $ \Omega \cdot cm $ were cleaned according to the standard procedure for silicon samples. After several hours of degassing, the samples were rapidly heated up to 1500K with direct current. The atomic chains were prepared by depositing about 0.44 ML (monolayer) of Au on a sample kept at room temperature and then annealing at about 870K for 5 minutes. In this article, 1 monolayer denotes the density of Si(111) surface atoms (7.84 $\times$ 10$^{14}$ atoms/cm$^{2}$). 

%

\begin{figure}[!htbp]
	\centering
	\includegraphics[width=1\columnwidth]{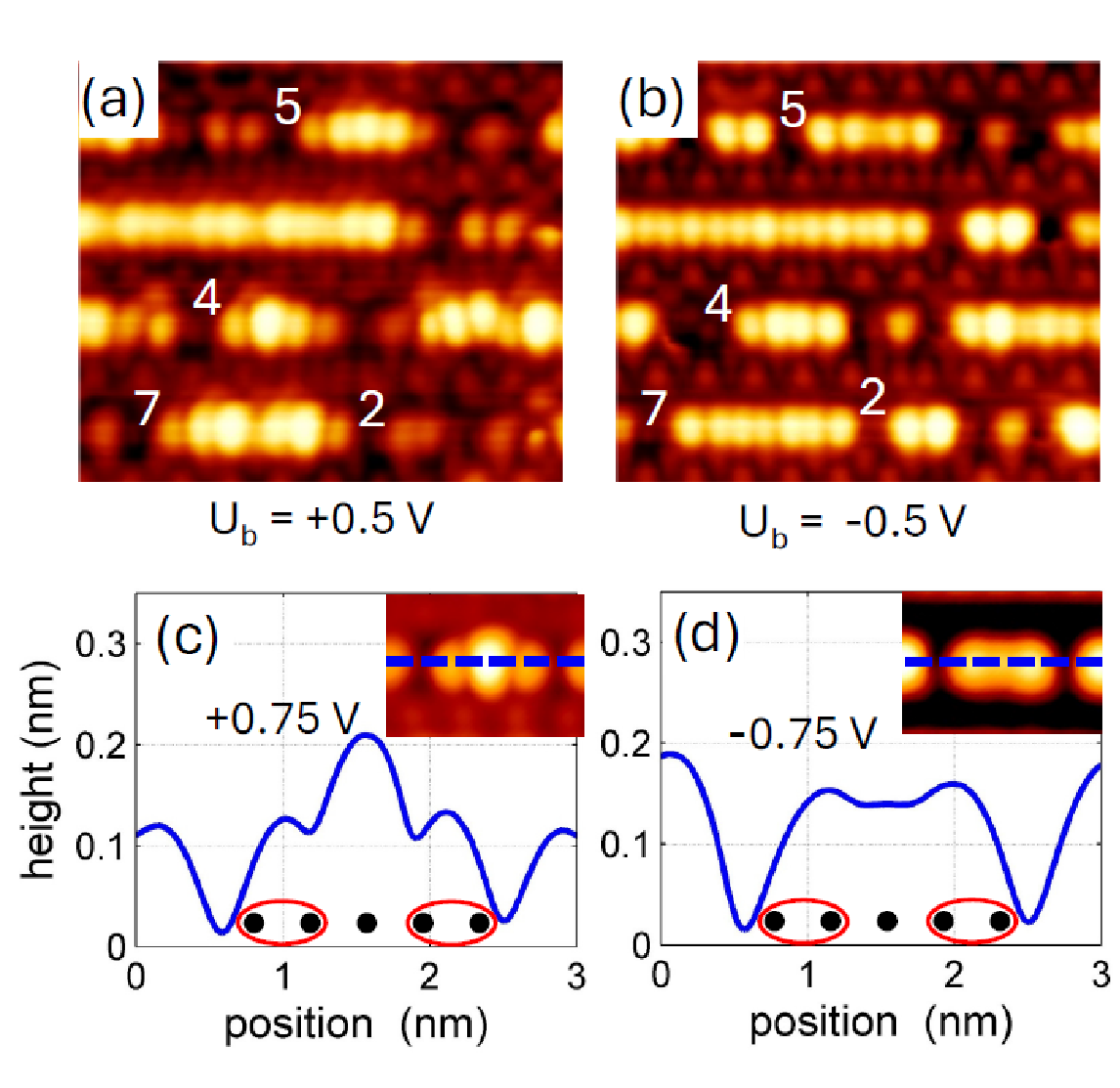}
	\caption{STM images of short Si atomic chains on Si(553)-Au surface at 110K and their topography variation upon reversing of sample bias polarity. Panels (a) and (b):  $7.7 nm \times 6.7 nm$ constant current images of the same sample area recorded with $I_{T} = 50 pA$ and sample bias of $0.5 V$ and $-0.5 V$, respectively. Numbers denote the effective length of chains, $N$, selected for further  analysis.
		Panels (c) and (d): 
		Topographic images of different Si(553)-Au sample at 77K and height profiles along  length of the same chain composed of $5$ Si atoms ($N = 3$ sites). Pairs of outer atoms encircles with red ellipses form dimers, which are treated as single sites. The images were recorded with sample bias of $0.75 V$ and $-0.75 V$, respectively. Tunneling current was set at $I_{T} = 50 pA$.  	 
	}
	\label{fig:stm1}
\end{figure}
%
Figure \ref{fig:stm1} (a), (b), show high resolution STM filled and empty state topographies taken at the same sample area.  Images were measured with a sample bias of $-0.5V$ and $+0.5$V, respectively. 
	Both images show alternately distributed bright Si edge atoms with unsaturated bonds and 
	double gold chains incorporated into the middle of each 1.45 nm wide Si(111) terrace \cite{Krawiec2010,Braun2018} (see also Fig.~\ref{fig01t}b).
The periodicity of the Au double chain is  equal to $2\times a_{Si[1\bar{1}0]}$. 
Owing to its highly regular structure, this chain was used as a reference to determine the number of atoms in the shorter, irregular Si chains.     
It can be seen that the modulation of the height of the structures depends on their length and on the polarity. 
%
Upon reversing the bias from $-0.5$ V to $+0.5$ V, some parts of the short chains change their intensity and appear as dark "blobs." It is well known  \cite{Yeom2022,Krawiec2010,Braun2018}  that these "blobs" represent pairs of coupled Si atoms with onsite wavefunctions that strongly overlap.
 Therefore  short chains depicted in Fig.~\ref{fig:stm1} 
can be identified as a structure composed of a finite effective number of sites, $N$, which are composed of alternating dimer sites and single Si sites.
Such restructuring of short Si chains is illustrated in Fig. \ref{fig:stm1} (c) and (d) 
on another Si(553)-Au substrate. 
Here insets show STM topographic images of the same chain recorded at sample bias equal to  $0.75$V and $-0.75$V, respectively. The chain consists of five Si atoms displaying features of dimerization characteristic of longer Si edge chains. In this configuration, the outer pairs of atoms form strongly bonded dimers, which are only weakly coupled to the central atom - a feature clearly illustrated by the height profiles shown in panels (c) and (d), as well as in  Fig.~\ref{fig01t}(b).

\section{\label{sec3} Theoretical description and analytical results}

\begin{figure}[!htbp]
	\begin{center}
		\includegraphics[width=7cm,keepaspectratio,]{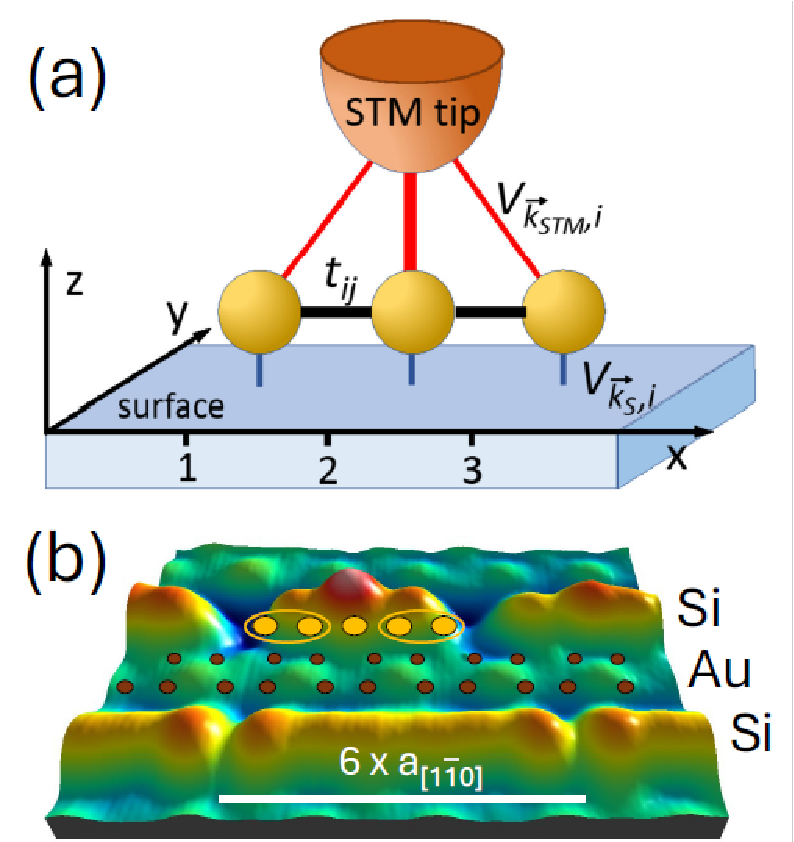}
	\end{center}
	\caption{Panel (a): Scheme of the STM system considered through work with the Cartesian coordinates $x,y$ and $z$. Chain sites are represented by golden spheres along $x$ axis with the electron hoppings between the neighbouring sites, $t_{ij}$ (black lines). The red lines correspond to the STM-chain couplings, $V_{\vec{k}_{STM},i}$, which values depend on the STM position over the surface, $V_{\vec{k}_{S},i}$ denotes chain-surface coupling.
		Panel (b): STM topographic 3D image of the Si(553)-Au surface with two Si step-edge chains, recorded at a tunneling current of $50pA$, and a sample bias of $0.25V$ at $77K$. In the middle of the terrace, the positions of the double Au chains are indicated by small brown ellipses. The atomic positions of Si atoms in the short step-edge chain are marked with yellow full ellipses. Although there are five Si atoms in this segment, due to dimerization at the edges, where outer atoms strongly hybridize, it is reasonable to consider the system as an effective 1D chain with $N=3$ sites.   
	}   \label{fig01t}
\end{figure}

To analyze the role of decoupled states in the STM topography of atomic systems, we calculated the electrical properties of short linear systems. The theoretical model under consideration consists of a few atomic sites 
(typically containing fewer than 7 atomic sites) arranged linearly on a substrate (acting as an electron reservoir), as schematically shown in Fig.~\ref{fig01t}(a). 
The STM tip represents an additional external electrode serving as an electrical probe. The gray spheres depict an atomic chain of length $N=3$ sites (which may correspond to either a single Si atom or a Si dimer), although other arbitrary chain lengths are also considered in the calculations. The black and red lines correspond to possible electron pathways, representing hopping integrals (couplings between the relevant sites or electrodes).
These elements between sites in the chain are constant, while the STM–chain couplings (red lines) depend in the relative distance between the tip and each atomic site in the chain.
Additionally, the atoms are coupled to the substrate, which serves as a bias electrode.
%
In panel (b) the corresponding STM topographic image of a single terrace between two Si step-edge chains (horizontal 1D structures) is shown. In the middle of the terrace, double Au chains are visible and marked with brown ellipses. The upper Si chain includes a short segment composed of five Si atoms (indicated by yellow ellipses); however, the terminal atoms strongly hybridise and form dimers. As a result, the effective length of the chain is reduced to $N=3$ sites, as schematically illustrated in panel (a).

The tight-binding Hamiltonian for the system depicted in Fig.~\ref{fig01t}  is expressed in the standard second-quantized  notation  and comprises two terms, $H=H_0+H_{int}$, where  $H_0$ corresponds to the single-electron component and $H_{int}$ describes the interaction term (between electron states), i.e.:
\begin{eqnarray} 	\label{001} 
H_{0}&=& \sum_{i=1}^N \varepsilon_i c^{\dagger}_i c_i + \sum_{\vec{k}}  \varepsilon_{\vec{k}} c^{\dagger}_{\vec{k}} c_{\vec{k}} \,, \nonumber\\
H_{int} &=& \sum_{i=1}^N\sum_{\vec{k}} V_{\vec{k} i} c^{\dagger}_{\vec{k}} c_{i} + \sum_{i=1}^{N-1} t_{i,i+1} c^{\dagger}_i c_{i+1} + H.c. \,, 
\end{eqnarray}
where $\vec{k}=\vec{k}_S, \vec{k}_{STM}$ are the electron wave vectors of the surface and STM leads,  and $c^{\dagger}_{\alpha}$, $c_{\alpha}$ are the electron creation and annihilation operators at the $\alpha$ site. Here we consider the spatial dependence of the STM coupling, $V_{\vec ki}(x,y,z) = V_{\vec ki} e^{-b \, \left( r_i(x,y,z)-1 \right)}$, 
 where the position of the $i$-th atomic site in the chain is given by $[x_i,y_i,z_i=0]$ and the relative distance between the STM tip and the chain site is  
 $r_i(x,y,z) = \sqrt{(x_i-x)^2+(y_i-y)^2+z^2}$. The $b$ parameter is a spatial decay coefficient (it governs the decay rate of the wavefunction overlap between the tip and a surface atom as a function in space). 
Thus each chain site is coupled with the STM electrode via the energy independent  effective  coupling which in the wide band approximation reads:
$\Gamma^{STM}_{ij}(x,y,z)= \Gamma^{STM}_{ij} e^{-2 \, b \, \left(r_i(x,y,z)-1\right)} \delta_{ij} $.
This approximation assumes that the leads (electronic reservoirs) exhibit a flat or almost energy-independent DOS around the Fermi level, without band gaps or sharp band edges — conditions that are satisfied in our experiments.
We model the  chain--surface coupling,  ${\Gamma}^{S}_{ij}=2 \pi \sum_{\vec  k_{S}} V_{\vec k_{S} i} V^*_{	\vec k_{S} j}\delta(E-\varepsilon_{\vec k_{S}})$, within the limit of localized electron states in the substrate, $k_F a \to \infty$ (large value of the surface  Fermi vector and lattice constant product), which within a wide band approximation  can be written in the following short form ${\Gamma}^{S}_{ij} =\Gamma^S \delta_{ij}$. In such a case each atomic site is effectively coupled with an individual electrode, so that an electron tunneling from the site to the surface can reappear only at the same atomic site. This assumption is quite reasonable for the  surface considered in our experiments.   
The electron hoppings between the chain sites are denoted by $t_{ij}$, and are considered within the nearest-neighbour approximation. This approach is justified for linear systems, where the strongest wavefunction overlap occurs between adjacent sites, and contributions from more distant atoms are negligible.

Electron transport properties through the system are analyzed within the framework of the Green's function method. 
The local DOS for each site is obtained from the relation $DOS_i(E)= -{1 \over \pi} Im G^ r_{ii}(E)$, where $G^r_{ii}(E)$ is a retarded Green function related to the $i$-th site of the chain and can be obtained from the equation of motion technique \cite{a09,a10}.
The knowledge of the Green's function allows one to obtain the transmission function, $T(E)=Tr\{\hat{\Gamma}^{STM} \hat{G}^r \hat{\Gamma}^{S} \hat{G}^a\}$,  and thus the STM current, $I={2e\over h}\int dE 
[f_{STM}(E)-f_{S}(E+V)] T(E) $, where $\hat{G}^{r,a}$ are the retarded and advanced Green
function matrices, $\hat{G}^a_{ij}=(\hat{G}^{r})^*_{ij}$,  $f(E)$ is the Fermi function of the tip or the surface electrode, and the external voltage is expressed by means of chemical potentials of both electrodes, $V =(\mu_{STM} - \mu_{S})/e$.

The above theoretical description can be applied to atomic chains of the length $N$ between the surface and STM electrodes. For such a regular chain, i.e. for the same hopping integrals along the chain, $t_{i,i+1}=t_1$, and for uniform on-site energies $\varepsilon_i=\varepsilon_0$  analytical calculations can be carried out. Assuming a wide STM tip for $b=0$, after some tedious algebra, we have obtained  the general analytical relation for the retarded Green's function $G_{ii}(E)$ at a given chain site, $i$:  
\begin{widetext}
	%
\begin{eqnarray}\label{equ:N3a}
	G_{ii}^r(E) &=& 
   	\frac{	{\prod_{j_1=1}^{i-1} \left( E-\varepsilon_0+i{\Gamma^S+\Gamma^{STM} \over 2} -2 t_1 \cos {j_1\pi \over i}  \right) } \,\,	
		{\prod_{j_2=1}^{N-i} \left( E-\varepsilon_0+i{\Gamma^S+\Gamma^{STM} \over 2} -2 t_1 \cos {j_2\pi \over N-i+1}  \right) } }
	  {\prod_{j_3=1}^{N} \left( E-\varepsilon_0+i{\Gamma^S+\Gamma^{STM} \over 2} -2 t_1 \cos {j_3\pi \over N+1}  \right) } \,,
\end{eqnarray}
%
\end{widetext}
 where  $\prod_{j=1}^{0}(...)=1$. Note that the system's eigenenergies correspond to the minima of the denominator in Eq.~\ref{equ:N3a}, which represents a product of $N$ independent terms. Consequently, there should generally be $N$ different solutions for the eigenstates. 
However, for certain values of the numerator's product, the total number of possible states decreases. This leads to a system eigenstate composed of a smaller number of localized atomic states, resulting in a decoupled state. As a consequence, this situation must be reflected in the local DOS structure, manifesting as a reduced number of local DOS peaks, which will be analyzed in Fig.~\ref{fig02t}.

The above expression for the retarded Green's function in an analytical form for a chain of arbitrary length on a surface is highly significant, as it enables the calculation of all relevant electrical properties, including the STM tunneling current. Although the above formula is somewhat complex, it  allows us to draw important conclusions regarding states decoupled from the surface.
One of the most important findings is that for an odd number of atoms in the chain  such decoupled states always exist in the system. In this case, assuming an odd $N$ ($N=2M+1$, where $M=1,2,...$), 
the Green's function associated with the central atom of the chain, $i=M+1=(N+1)/2$, reads
%
\begin{eqnarray}\label{equ:Nodd}
	G_{ii}^r(E) &=& \frac{\prod_{j=1}^{M} \left(  E-\varepsilon_0+i{\Gamma^S+\Gamma^{STM} \over 2}-2 t_1 \cos {j\pi \over M+1}  \right)^2 }{\prod_{j=1}^{2M+1} \left(E-\varepsilon_0+i{\Gamma^S+\Gamma^{STM} \over 2} -2 t_1 \cos {{j \over 2}\pi \over M+1}  \right) } \,\nonumber\\ \end{eqnarray}
%
As a result, the local DOS at this middle site exhibits only $(N+1)/2$ peaks (instead of $N$ states), meaning that at least one atomic state in the system is decoupled from the external electrodes. This implies that for a linear chain of length  $N=3,5,7,...$ such decoupled states always exist. On the other hand, for a chain of arbitrary length, the edge sites ($i=1$ or $i=N$) always exhibit the full set of $N$ eigenstates, meaning that no decoupled states are associated with these atoms. As a consequence, a system consisting of a single atom or two coupled atoms will never exhibit states decoupled from the surface. In the latter case, for the system isolated from the electrodes, the Hamiltonian can be diagonalized, and both molecular eigenstates can be expressed in terms of the individual atomic states. This confirms the analysis performed using the analytical formulas for the retarded Green's function.

\section{\label{sec3b}Atomic chains with decoupled states}

In the following we analyze atomic chains composed of a few atoms only. 
The presence of decoupled states not only determines the form of the eigenfunctions but also affects the local DOS and differential conductance characteristics. As a result, signatures of these states can be experimentally observed in atomic systems.
As shown in the previous section, decoupled states do not appear in a single-atom system or a two-atom system. However, the situation changes for three linearly arranged sites, $N=3$, coupled via the nearest-neighbour hopping integrals $t_1$, which corresponds to the geometry of Refs. \onlinecite{Poltl2009,Michaelis2006}.
In this case, the system has three eigenstates:
$\Psi_{1}={1 \over 2} (\phi_1 - \sqrt{2} \phi_2 +\phi_3)$, $\Psi_{2}={1 \over \sqrt{2}} (\phi_1 -\phi_3)$, and $\Psi_{3}={1 \over 2} (\phi_1 + \sqrt{2} \phi_2 +\phi_3)$, 	
with eigenvalues:  $E_1=\varepsilon_0-\sqrt{2} t_1$, $E_2=\varepsilon_0$, and $E_3=\varepsilon_0 + \sqrt{2} t_1$, respectively ($\phi_i$ are localized atomic states). The molecular states $\Psi_{1}$ and $\Psi_{3}$ are composed of all atomic states but  the second molecular state, $\Psi_{2}$, does not include the middle-atom state, $\phi_2$, and this state is dark  in the sense that it is decoupled from the environment (and no electrons can flow through the middle atom) \cite{Weymann2011}. 
%

%
\begin{figure}[!htbp]
	\begin{center}
		\includegraphics[width=8cm,keepaspectratio,]{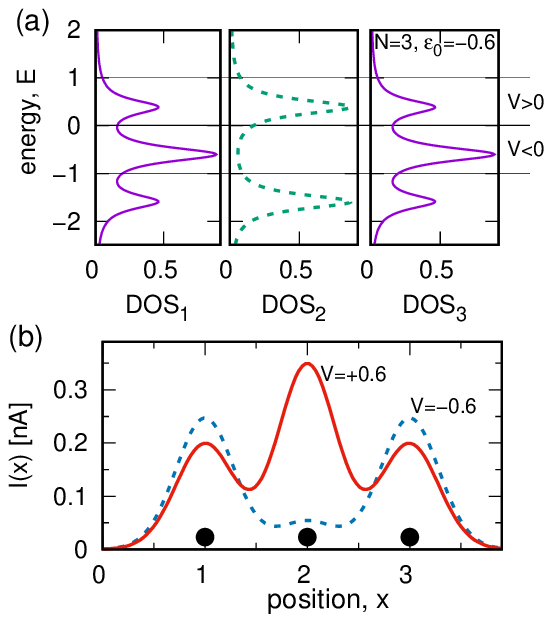}
	\end{center}
	\caption{Panel (a) - Local DOS at each site for the chain length, $N=3$. Panel (b) shows the calculated STM current profiles at constant tip height along the chain (positions of sites are $x=1,2,3$, $y=1$, $z=0$) for positive and negative biases,  $V=\pm 0.6$V, and for $\varepsilon_i=-0.6$eV, $t_1=0.7$eV, $\Gamma=0.35$eV,  $b=10$, and the tip-sample distance $1.6$ (in units of the interatomic distance). Black dots in the bottom panel symbolize the positions of atomic sites (single atoms or dimers).}
	\label{fig02t}
\end{figure}
%
The form of the system's eigenfunctions is reflected in its electronic properties. To begin, we will determine the STM current for this system using the analytical expression for the retarded Green's function, given in Eq.~\ref{equ:N3a}.
Taking the imaginary part of $G^r_{ii}$ we can find the spectral density function at each site (DOS$_i$). The corresponding local DOS curves are depicted in the upper panel in Fig.~\ref{fig02t}. 
Note that in our calculations, the tight-binding (TB) effective parameters were adjusted to reflect the characteristics of real atomic chains and were  optimized to achieve a satisfactory agreement with the experimental observations.	
Here, the two edge sites of the chain, $i=1$ and $i=3$ (solid lines for DOS$_1$ and DOS$_3$), are characterized by a strong central peak and two smaller side peaks in the local DOS, with eigenenergies $E_{1/2/3}$, but the middle site of the chain ($i=2$) exhibits only two side peaks (dashed curve). The absence of a central peak in the local DOS at this site is a direct consequence of the structure of $\Psi_2$, which consists only of the local states $\phi_1$ and $\phi_3$.  
Thus, the existence of decoupled states in the system is manifested at the middle site of the chain by two side peaks in the local DOS function, with a local minimum at $E=\varepsilon_0$. 
The presence of decoupled states in the local DOS should also be reflected in the current-voltage characteristics and STM topography.  
To illustrate this prediction we show in Fig.~\ref{fig02t}(b) the calculated  STM current profiles, $I(x)$, for fixed tip-sample distance (equal to 1.6 units of the interatomic distance) along the chain composed of $N=3$ sites. We show the results for  positive and negative STM voltages, $V=\pm 0.6$V. As one can see for positive bias (solid red line) the STM current reproduces all positions of sites (each site has a local DOS peak inside the energy voltage window). For the middle site the value of both DOS peaks is larger than for the end sites which is also reflected in the current values through the chain (solid line). However, for the negative bias (dashed curve in panel b) the current reveals different behaviour. We observe large current values for the end sites ($x=1$ and $x=3$) but the current  flowing to the second site is much smaller  due to the local minimum of DOS for $E=\varepsilon_0$. The lack of a local DOS peak for this energy blocks the STM current. 
Thus, the vanishing tunneling current observed at the center of the chain under negative bias is a consequence of the presence of decoupled states in the system. This significantly alters the topographic images for both STM polarizations, to the extent that a single atom may become invisible to the STM tip.
Note that although the profiles in Fig.~\Ref{fig02t}(b) display the variation of the tunneling current at a fixed $z$-height, there is a close relationship to the constant-current profile representation, especially when the same atoms form a linear structure - both profile types measure the distribution of spatial local DOS.
This is exactly what we observe in the STM topography shown in Fig.~\ref{fig:stm1}, namely the profile lines for both polarizations in panels (c) and (d). 
This constitutes a key result of our work: The study demonstrates that decoupled states in stable atomic structures on a surface can be successfully investigated using STM techniques, and crucially, the presence of these states can drastically suppress the tunneling current.

It is worth noting that similar atomic chains on the Si(553)-Au substrate have been considered previously in terms of the occurrence of end states in these chains.\cite{Crain2005} The authors found that under positive bias, the end atoms are barely visible, whereas under negative bias, these atoms are significantly enhanced — an effect we also observe in our studies. However, explaining their results within the framework of the TB model required changes and diversification of the on-site (electron binding) energies, despite considering the regular Si chains.
This assumptions leads to a transfer of DOS from the empty to the filled states which may explain the end state bahavior but cannot generate the physics of decoupled states. 
Such a small redistribution of the end-atom energies can manifest as slightly shifted topographic maxima at the chain edges when the voltage polarity is reversed (which is also visible in our experimental data, panels c and d in Fig.~\ref{fig:stm1})
 In our model we concentrate on the dark state physics and consider only periodical and regular 1D chains thus it is reasonable to describe them by the same energy parameters. In such a case dark states can appear and they are responsible for variations of topography results due to the bias voltage.



\textbf{Longer chains $N>3$}.  
In this part we are going to discuss decoupled states in longer chains. It is interesting that for  a four-atom chain in a linear geometry, the system's eigenstates always consist of all four atomic states, meaning that, in general, no decoupled states exist in such systems. We have experimentally confirmed that an atomic chain composed of $N = 4$ sites behaves as predicted by theory.
%
A very interesting system for analyzing states decoupled from the surface is a linear chain with five sites where, due to odd $N$, at least one decoupled state always exists. Such a system is described by five eigenstates, $\Psi_{i}$, with eigenenegies given by $E_i=\varepsilon_0+2t\cos \frac{i\pi}{6}$, ($i=1,...,5$). 
In particular for $N=5$ only two of the system eigenfunctions  are expressed by all 5 atomic states, $\Psi_{1}$ and $\Psi_{5}$, while the remaining three,  $\Psi_{2}, \Psi_{3}, \Psi_{4}$,  consist only of three or four states, e.g. $\Psi_{3}={1 \over \sqrt{3}} \left(\phi_1-\phi_3+\phi_5\right)$. Therefore, there are atomic states that are decoupled from the surface. 
%
\begin{figure}[!htbp]
	\centering
	\includegraphics[width=8.0cm,keepaspectratio,]{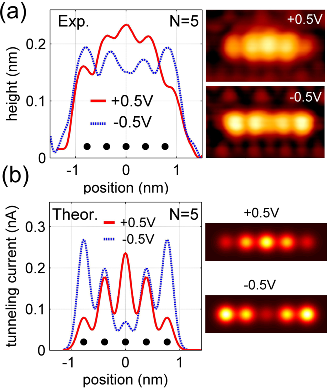}
	\caption{Panel (a) - STM topography images registered at biases of +0.5V and -0.5V with tunneling current of 50 pA and corresponding height profiles along the chain composed of 5 sites (N = 5). Panel (b) - Calculated tunneling current images and profiles along chain from panel (a). The other parameters are: $\varepsilon_0=-0.6$eV, $\Gamma=0.35$eV, $t_{12}=t_{45}=1.2$, $t_{23}=t_{34}=2$, $b=10$, $N=5$, and bias voltage $\pm 0.5$V. }
	\label{fig:stm2}
\end{figure}
%

Figure~\ref{fig:stm2} shows the experimental height profiles of a linear chain composed of  $N=5$ sites for negative and positive bias along with their topographic images (panel a).
For positive STM voltage the number of peaks corresponds to number of sites,  whereas for a negative bias, only four peaks are observed. The absence of a local DOS peak at the central atom indicates the formation of decoupled states.
As a result, if the STM tip is positioned above a chain atom hosting such a decoupled state (and its energy falls within the STM bias window), that site remains nearly invisible to the tip. 

The calculated tunneling current for the considered atomic chain is shown in Fig.~\ref{fig:stm2}, panel (b), for both positive and negative biases. It is evident that for both STM polarizations, a different number of peaks appear in the current (or height) profiles in this case. 
The agreement between the theoretical and experimental results in Fig.\ref{fig:stm2} in panels (a) and (b) is quite satisfactory, confirming that the voltage dependence of the STM topography images can be explained by the asymmetrical structure of the local DOS with decoupled states.
The analysis conducted for longer chains, such as for $N=7$, also reveals the existence of decoupled states in these linear systems. Nevertheless, owing to the increased number of molecular states and their mutual overlap, the effect appears less pronounced. 
%


\section{\label{sec4} Time Dynamics of Decoupled States }

In this section, we investigate the time dynamics of a system featuring surface-decoupled states.
Time evolution characteristics allow us to gain insight into the system’s dynamics, including the rate at which it relaxes to its equilibrium state.
To this end, we analyze a simple atomic chain subjected to a sudden change in inter-site couplings, effectively transforming it into a system that hosts decoupled states. This analysis reveals whether these states emerge instantaneously or require a finite time to develop, during which nonequilibrium effects gradually diminish. Consequently, in the considered case, the system’s Hamiltonian (Eq.~\ref{001}) becomes time-dependent through the coupling parameters between the system’s states, denoted as $t_{ij}(t)$. 
The time dynamics of the system are calculated within the interaction picture using the evolution operator method \cite{Grim,Tar2003,Zhou2008,Kwap2020}, where the equation of motion takes the form ($\hbar=1$):
%
\begin{equation}\label{equ:equ1}
	i \frac{\partial }{\partial t} U(t, t_0) = \hat{V}(t) U(t, t_0) \, ,
\end{equation}
where $\hat{V}(t)=U_0(t,t_0) H_{int}(t) U^\dagger (t,t_0)$,  $U_0(t,t_0)=\mathcal{T} \exp{\left( i\int_{t_0}^t
	dt' H_0(t') \right)}$ and $\mathcal{T}$ is the time ordering operator. The electronic properties of the system are expressed by the evolution operator matrix elements obtained from Equation~(\ref{equ:equ1}).
In our case the atom--surface elements of the evolution operator take the following differential equation:
\begin{eqnarray}  \label{equ:Uik_differential2}
	{d U_{i,k}(t) \over d t}  &=& -i \sum_{i'} t_{ii'}(t)e^{i(\varepsilon_{i'}-\varepsilon_i)t} U_{i', \vec{k}}(t)  \nonumber\\ &-& i {V}_{i,\vec{k}}(t) e^{i(\varepsilon_i-\varepsilon_{\vec{k}})\, t} \\ 
	&-&  V_{i,k}^2 \int_{0}^t dt' D(t-t') e^{i \varepsilon_i (t-t') }  
	U_{i,k}(t')  \nonumber \,,
\end{eqnarray}
where we put $t_0=0$, thus $U_{i,k}(t,t_0)= U_{i,k}(t)$,  and  $D(t-t')=\int dE
D_{\alpha}(E) \exp(-i E (t-t'))$, is the Fourier transform of the electron
DOS in the $\alpha$ electrode.
%
The time dependent spectral density function at each site in the chain satisfies the relation  \cite{Kwap2020}
\begin{eqnarray}
	DOS_i(E,t)=\sum_{\alpha} D_{\alpha}(E) | U_{i, \alpha}(E,t) |^2 \,, \label{a03}
\end{eqnarray}
Thus, knowledge of the $U_{i,k}(t)$ elements, which we compute numerically from Eq.~(\ref{equ:Uik_differential2}), is essential for determining the time-dependent local DOS under arbitrary time variations of the coupling parameters.

It is worth noting that in some special cases (regular and symmetric chains with identical on-site energies and uniform couplings), the set of differential equations for $U_{i,k}(t)$ can be solved analytically using the Laplace transform technique \cite{Kwap2020,Kwap2023}, which allows us to study initial transient effects. However, in our case, the Hamiltonian parameters evolve at longer times, even after the system reaches equilibrium conditions. As a result, we primarily rely on numerical evaluation of the  $U_{i,k}(t)$ elements assuming the wide-band approximation for the surface DOS. This approximation slightly simplifies the calculations, as the last term in Eq.~\ref{equ:Uik_differential2} then takes the form $-\Gamma/2 \,\, U_{i,k}(t)$.
In these calculations, we consider the zero-temperature case, and the energy and time unit are given by $\Gamma$ and  $\hbar / \Gamma$, respectively, thus, for $\Gamma$ on the order of meV, the time units are on the order of $ps$, and  the reference energy is set to the Fermi energy of the surface electrode, $E_F = 0$.
Note that the STM technique is not suitable for probing ultrafast time dynamics on the nanosecond or picosecond scale. For experimental studies of such temporal phenomena, more sophisticated optical methods (such as time-resolved photoemission spectroscopy) are more appropriate.


%
\begin{figure}[!htbp]
	\centering
	\includegraphics[width=8.7cm,keepaspectratio,]{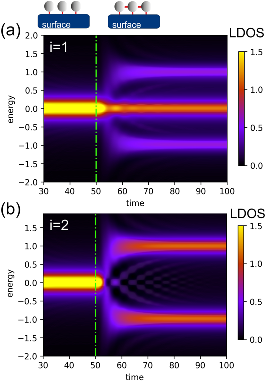}
	\caption{Local DOS as a function of time and energy for the system composed from 3 sites at the surface  schematically shown in the upper diagrams. The upper panel corresponds to the local DOS at the end sites, $i=1,3$, and the bottom panel at the middle site, $i=2$. At time $t=50$ the couplings between sites are switched on from $t_{12}=t_{23}=t_i=0$ to $t_i=0.7$.  The other parameters are: $\varepsilon_{1/2/3}=\varepsilon_0=0$eV, $\Gamma=0.35$eV. }
	\label{fig:3Da}
\end{figure}
%
The minimal system in which decoupled states can be observed consists of three sites. Therefore, we begin by analyzing the time evolution of the local DOS function for such a system which is schematically shown in the upper diagrams in Fig.~\ref{fig:3Da}.
This figure shows the time and energy dependence of the local  DOS function associated with individual sites (top panel – sites 1 and 3 which are the same due to the spatial symmetry; bottom panel – site 2). Bright regions correspond to high DOS values. As seen in the initial stage, $t<50$, these atoms do not interact with each other, so each site is characterized by a time-independent DOS function with a global maximum at $E=\varepsilon_0=0$. 
At $t=50$, the couplings between all atoms are switched on, $t_i=0.7$,  causing the electronic structure of the system to reorganize. We observe that on the edge atoms, $i=1,3$, the main DOS peak decreases in intensity, while two new side peaks emerge at $E=\varepsilon_0 \pm t_i$. Thus, for large times, we clearly see three distinct peaks in the local DOS structure, corresponding to the three molecular states of the system. The formation of these states occurs relatively quickly after the sudden activation of couplings, within approximately $25-30$ time units.
Similar side peaks in the DOS appear over time for the central atom ($i=2$, bottom panel). However, in this case, the central DOS peak gradually disappears, so at large times, only two LDOS peaks remain. This is because one of the atomic states (the state of the central atom) becomes decoupled from the electrode. 
This energy region, where the main DOS peak vanishes over time, carries key information about how quickly a decoupled state forms in the system. Therefore, it requires a more detailed analysis.

\begin{figure}[!htbp]
	\centering
	\includegraphics[width=7.5cm,keepaspectratio,]{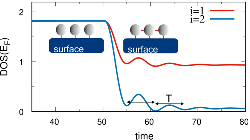}
	\caption{Time dependent local DOS at the Fermi energy for the system discussed in  Fig.~\ref{fig:3Da}. The red (blue) curve corresponds to the site $i=1$ ($i=2$), and all parameters are the same as in Fig.~\ref{fig:3Da}.}
	\label{fig:2D}
\end{figure}
In Fig.~\ref{fig:2D} we analyze the time evolution of the local DOS function at the Fermi energy, $E=E_F=0$ for $i=1$ and for $i=2$. This energy corresponds to the eigenenergy of the central molecular state in the coupled three-atom system. 
Upon an abrupt change in the couplings between atoms, the DOS function for the end sites (red line) decreases non-monotonically, exhibiting characteristic oscillations. The system approaches a new equilibrium state after approximately 25 time units. For the central atom (blue line), the DOS at the Fermi level drops sharply to almost zero, followed by several transient local maxima in the curve over time. These maxima indicate a non-equilibrium regime in the system, where a decoupled state is gradually forming.
The oscillations visible in this plot are associated with dynamic processes of local DOS structure formation across all atoms. At long timescales, this function converges to the local DOS profile discussed in  Fig.~\ref{fig02t}. Similar DOS oscillations have been observed in non-equilibrium processes during the analysis of transient crystal phenomena \cite{Kwap2020,Kwap2023}. 
It can be demonstrated that for the system under study, the period of these oscillations is $T={2\pi \over \sqrt{2}  t_i} \sim 6.3$. These oscillations are related to Rabi-like dynamics in atomic systems, originating from quantum interference between the collective eigenstates of the chain, and where the amplitudes of the oscillations depend on the initial state preparation. Note, that the energy splitting between the molecular states in our 3-atom system is $\sqrt{2} t_i$, and while the middle atomic state remains decoupled from the substrate (and thus does not contribute directly to the local DOS signal), its energy separation from the edge molecular states plays a critical role in sustaining the oscillation frequency (and period $T$).

\begin{figure}[!htbp]
	\centering
	\includegraphics[width=8.5cm,keepaspectratio,]{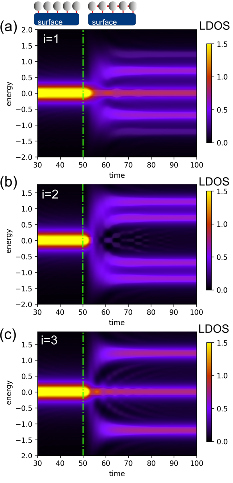}
	\caption{Local DOS as a function of time and energy for the system composed from $N=5$ (as shown in the upper diagrams). Panels (a), (b) and (c) show the local DOS at the chain sites $i=1$, $i=2$ and $i=3$, respectively. At time $t=50$ the couplings between all atoms are switched on from $t_i=0$ to $t_i=0.7$.  The other parameters are: $\varepsilon_{i}=\varepsilon_0=0$eV, $\Gamma=0.35$eV.  }
	\label{fig:3Db}
\end{figure}
%
Now we consider longer chains and in Fig.~\ref{fig:3Db}, we analyze the time-dependent evolution of the local DOS function  in a five-site chain. As in previous studies, the abrupt activation of interatomic couplings occurs at time $t=50$. The outermost sites ($i=1$, which is the same as $i=5$, upper panel) exhibit contributions from all five molecular states in the system, as evidenced by five distinct DOS peaks for large time. The time evolution of DOS in this case is similar to that observed for edge sites discussed earlier for $N=3$: the main peak decreases in intensity in an oscillatory manner over time, and four new DOS peaks emerge at energies corresponding to the system's eigenenergies. In contrast, sites $i=2$ and 4 (middle panel) display only four local DOS peaks at sufficiently long timescales, indicating that one molecular state lacks any spectral weight associated with the central atom (no DOS peak at energy $E=0$). Similarly, the central atom (bottom panel) is characterized by only three molecular states at long times, implying that two atomic states remain decoupled from the substrate.
Notably, in both cases involving decoupled ("dark") states, the transient dynamics of LDOS formation exhibit characteristic decaying oscillations, linked also to Rabi-like processes arising from quantum interference between collective eigenstates.  
As illustrated in this figure, the duration of DOS oscillations -- and thus the time required to establish a substrate-decoupled state -- persists for approximately 25-30 time units in the system (the same period was observed also for longer chains $N=6$ and $N=7$). This timescale closely matches that observed in the 3-atom configuration, suggesting a universal relaxation mechanism governed by the coupling strength and system symmetry. Beyond $t>80$, the system reaches equilibrium, marked by the stabilization of the local DOS and the absence of transient oscillations.
The similarity in equilibration times between the 3-atom and longer chains implies that the effective decoupling dynamics are dominated by local nearest-neighbor interactions rather than the total chain length. This result is consistent with the STM topography images of Si chains, where atomic structures with a periodicity of three atoms are clearly visible also for longer chains. This indicates that the atomic arrangement within the chain is primarily influenced by nearest-neighbor interactions along the chain and the chain-substrate couplings.

\section{\label{sec5} Conclusions}
 
In this work, we have presented both theoretical and experimental investigations of atomic chains fabricated on the Si(553)-Au surface in the STM geometry, focusing on states that are decoupled from the substrate. We analyzed their role in electronic transport and STM topography.
Analytical expressions for the Green’s functions were derived, allowing us to calculate the local DOS and to identify decoupled states for chains of arbitrary length. We demonstrated that the presence of these states significantly affects the local DOS and the tunneling current. Notably, the number of peaks observed in the STM current (or height) profiles differs depending on the polarity of the applied voltage. This key result offers new insights into the interpretation of STM topographic images, as certain atomic sites may become entirely invisible to the tip due to suppressed current caused by the presence of decoupled states.

Moreover, we found that in chains with an odd number of atoms, a decoupled state naturally emerges, opening new possibilities for exploring such phenomena in more complex atomic structures. We also showed that these states can form in systems created via self-organization of atoms on vicinal surfaces, offering a route to fabricate arrays of decoupled states.
In contrast to previous theoretical studies of dark states in quantum dot systems - typically involving meV energy scales - we explored real atomic systems characterized by eV-scale energies, making them potentially more relevant for practical applications.
Finally, we demonstrated that decoupled states do not form instantaneously but rather emerge over a finite period of time. Time-resolved analysis of the local DOS reveals characteristic transient oscillations with distinct time-dependent patterns that gradually stabilize into equilibrium. This relaxation process, governed primarily by nearest-neighbor interactions and chain–surface coupling, appears to be largely independent of the chain length and represents a universal mechanism for the formation of such states.

These findings can serve as a foundation for exploring new theoretical models of electron localization and transport in low-dimensional systems. The identification of decoupled states in real atomic chains suggests opportunities for designing atomic-scale devices that exploit such states for functionality, such as quantum interference elements or switches where controllable decoupling is desirable. The ability to realize and manipulate decoupled states at room temperature and on accessible energy scales may also support the development of novel device architectures based on atomically precise surface engineering.


\textit{Acknowledgments}: This work was supported by National Science Centre, Poland, under Grant No. 2022/45/B/ST3/01123.  
	
\bibliography{bibliography}

\providecommand{\noopsort}[1]{}\providecommand{\singleletter}[1]{#1}%
\begin{thebibliography}{41}%
\makeatletter
\providecommand \@ifxundefined [1]{%
 \@ifx{#1\undefined}
}%
\providecommand \@ifnum [1]{%
 \ifnum #1\expandafter \@firstoftwo
 \else \expandafter \@secondoftwo
 \fi
}%
\providecommand \@ifx [1]{%
 \ifx #1\expandafter \@firstoftwo
 \else \expandafter \@secondoftwo
 \fi
}%
\providecommand \natexlab [1]{#1}%
\providecommand \enquote  [1]{``#1''}%
\providecommand \bibnamefont  [1]{#1}%
\providecommand \bibfnamefont [1]{#1}%
\providecommand \citenamefont [1]{#1}%
\providecommand \href@noop [0]{\@secondoftwo}%
\providecommand \href [0]{\begingroup \@sanitize@url \@href}%
\providecommand \@href[1]{\@@startlink{#1}\@@href}%
\providecommand \@@href[1]{\endgroup#1\@@endlink}%
\providecommand \@sanitize@url [0]{\catcode `\\12\catcode `\$12\catcode
  `\&12\catcode `\#12\catcode `\^12\catcode `\_12\catcode `\%12\relax}%
\providecommand \@@startlink[1]{}%
\providecommand \@@endlink[0]{}%
\providecommand \url  [0]{\begingroup\@sanitize@url \@url }%
\providecommand \@url [1]{\endgroup\@href {#1}{\urlprefix }}%
\providecommand \urlprefix  [0]{URL }%
\providecommand \Eprint [0]{\href }%
\providecommand \doibase [0]{https://doi.org/}%
\providecommand \selectlanguage [0]{\@gobble}%
\providecommand \bibinfo  [0]{\@secondoftwo}%
\providecommand \bibfield  [0]{\@secondoftwo}%
\providecommand \translation [1]{[#1]}%
\providecommand \BibitemOpen [0]{}%
\providecommand \bibitemStop [0]{}%
\providecommand \bibitemNoStop [0]{.\EOS\space}%
\providecommand \EOS [0]{\spacefactor3000\relax}%
\providecommand \BibitemShut  [1]{\csname bibitem#1\endcsname}%
\let\auto@bib@innerbib\@empty
\bibitem [{\citenamefont {Alzetta}\ \emph {et~al.}(1976)\citenamefont
  {Alzetta}, \citenamefont {Gozzini},\ and\ \citenamefont {Moi}}]{Alzetta1976}%
  \BibitemOpen
  \bibfield  {author} {\bibinfo {author} {\bibfnamefont {G.}~\bibnamefont
  {Alzetta}}, \bibinfo {author} {\bibfnamefont {A.}~\bibnamefont {Gozzini}},\
  and\ \bibinfo {author} {\bibfnamefont {G.}~\bibnamefont {Moi}, \bibfnamefont
  {L.and~Orriols}},\ }\bibfield  {title} {\bibinfo {title} {An experimental
  method for the observation of r.f. transitions and laser beat resonances in
  oriented na vapour},\ }\href {https://doi.org/10.1007/BF02749417} {\bibfield
  {journal} {\bibinfo  {journal} {Il Nuovo Cimento B}\ }\textbf {\bibinfo
  {volume} {36}},\ \bibinfo {pages} {5} (\bibinfo {year} {1976})}\BibitemShut
  {NoStop}%
\bibitem [{\citenamefont {Aharon}\ \emph {et~al.}(2016)\citenamefont {Aharon},
  \citenamefont {Pozner}, \citenamefont {Lifshitz},\ and\ \citenamefont
  {Peskin}}]{Aharon2016}%
  \BibitemOpen
  \bibfield  {author} {\bibinfo {author} {\bibfnamefont {E.}~\bibnamefont
  {Aharon}}, \bibinfo {author} {\bibfnamefont {R.}~\bibnamefont {Pozner}},
  \bibinfo {author} {\bibfnamefont {E.}~\bibnamefont {Lifshitz}},\ and\
  \bibinfo {author} {\bibfnamefont {U.}~\bibnamefont {Peskin}},\ }\bibfield
  {title} {\bibinfo {title} {Multi-bit dark state memory: Double quantum dot as
  an electronic quantum memory},\ }\href {https://doi.org/10.1063/1.4972340}
  {\bibfield  {journal} {\bibinfo  {journal} {Journal of Applied Physics}\
  }\textbf {\bibinfo {volume} {120}},\ \bibinfo {pages} {244301} (\bibinfo
  {year} {2016})}\BibitemShut {NoStop}%
\bibitem [{\citenamefont {Fleischhauer}\ and\ \citenamefont
  {Lukin}(2000)}]{Fle2000}%
  \BibitemOpen
  \bibfield  {author} {\bibinfo {author} {\bibfnamefont {M.}~\bibnamefont
  {Fleischhauer}}\ and\ \bibinfo {author} {\bibfnamefont {M.~D.}\ \bibnamefont
  {Lukin}},\ }\bibfield  {title} {\bibinfo {title} {Dark-state polaritons in
  electromagnetically induced transparency},\ }\href
  {https://doi.org/10.1103/PhysRevLett.84.5094} {\bibfield  {journal} {\bibinfo
   {journal} {Phys. Rev. Lett.}\ }\textbf {\bibinfo {volume} {84}},\ \bibinfo
  {pages} {5094} (\bibinfo {year} {2000})}\BibitemShut {NoStop}%
\bibitem [{\citenamefont {An}\ \emph {et~al.}(2022)\citenamefont {An},
  \citenamefont {Kohno}, \citenamefont {Litvinenko}, \citenamefont {Seeger},
  \citenamefont {Naletov}, \citenamefont {Vila}, \citenamefont {de~Loubens},
  \citenamefont {Ben~Youssef}, \citenamefont {Vukadinovic}, \citenamefont
  {Bauer}, \citenamefont {Slavin}, \citenamefont {Tiberkevich},\ and\
  \citenamefont {Klein}}]{An2022}%
  \BibitemOpen
  \bibfield  {author} {\bibinfo {author} {\bibfnamefont {K.}~\bibnamefont
  {An}}, \bibinfo {author} {\bibfnamefont {R.}~\bibnamefont {Kohno}}, \bibinfo
  {author} {\bibfnamefont {A.~N.}\ \bibnamefont {Litvinenko}}, \bibinfo
  {author} {\bibfnamefont {R.~L.}\ \bibnamefont {Seeger}}, \bibinfo {author}
  {\bibfnamefont {V.~V.}\ \bibnamefont {Naletov}}, \bibinfo {author}
  {\bibfnamefont {L.}~\bibnamefont {Vila}}, \bibinfo {author} {\bibfnamefont
  {G.}~\bibnamefont {de~Loubens}}, \bibinfo {author} {\bibfnamefont
  {J.}~\bibnamefont {Ben~Youssef}}, \bibinfo {author} {\bibfnamefont
  {N.}~\bibnamefont {Vukadinovic}}, \bibinfo {author} {\bibfnamefont
  {G.~E.~W.}\ \bibnamefont {Bauer}}, \bibinfo {author} {\bibfnamefont {A.~N.}\
  \bibnamefont {Slavin}}, \bibinfo {author} {\bibfnamefont {V.~S.}\
  \bibnamefont {Tiberkevich}},\ and\ \bibinfo {author} {\bibfnamefont
  {O.}~\bibnamefont {Klein}},\ }\bibfield  {title} {\bibinfo {title} {Bright
  and dark states of two distant macrospins strongly coupled by phonons},\
  }\href {https://doi.org/10.1103/PhysRevX.12.011060} {\bibfield  {journal}
  {\bibinfo  {journal} {Phys. Rev. X}\ }\textbf {\bibinfo {volume} {12}},\
  \bibinfo {pages} {011060} (\bibinfo {year} {2022})}\BibitemShut {NoStop}%
\bibitem [{\citenamefont {Grynberg}\ and\ \citenamefont
  {Robilliard}(2001)}]{Gry2001}%
  \BibitemOpen
  \bibfield  {author} {\bibinfo {author} {\bibfnamefont {G.}~\bibnamefont
  {Grynberg}}\ and\ \bibinfo {author} {\bibfnamefont {C.}~\bibnamefont
  {Robilliard}},\ }\bibfield  {title} {\bibinfo {title} {Cold atoms in
  dissipative optical lattices},\ }\href
  {https://doi.org/https://doi.org/10.1016/S0370-1573(01)00017-5} {\bibfield
  {journal} {\bibinfo  {journal} {Physics Reports}\ }\textbf {\bibinfo {volume}
  {355}},\ \bibinfo {pages} {335} (\bibinfo {year} {2001})}\BibitemShut
  {NoStop}%
\bibitem [{\citenamefont {Michaelis}\ \emph {et~al.}(2006)\citenamefont
  {Michaelis}, \citenamefont {Emary},\ and\ \citenamefont
  {Beenakker}}]{Michaelis2006}%
  \BibitemOpen
  \bibfield  {author} {\bibinfo {author} {\bibfnamefont {B.}~\bibnamefont
  {Michaelis}}, \bibinfo {author} {\bibfnamefont {C.}~\bibnamefont {Emary}},\
  and\ \bibinfo {author} {\bibfnamefont {C.~W.~J.}\ \bibnamefont {Beenakker}},\
  }\bibfield  {title} {\bibinfo {title} {All-electronic coherent population
  trapping in quantum dots},\ }\href
  {https://doi.org/10.1209/epl/i2005-10458-6} {\bibfield  {journal} {\bibinfo
  {journal} {Europhysics Letters ({EPL})}\ }\textbf {\bibinfo {volume} {73}},\
  \bibinfo {pages} {677} (\bibinfo {year} {2006})}\BibitemShut {NoStop}%
\bibitem [{\citenamefont {Emary}(2007)}]{Emary2007}%
  \BibitemOpen
  \bibfield  {author} {\bibinfo {author} {\bibfnamefont {C.}~\bibnamefont
  {Emary}},\ }\bibfield  {title} {\bibinfo {title} {Dark states in the
  magnetotransport through triple quantum dots},\ }\href
  {https://doi.org/10.1103/PhysRevB.76.245319} {\bibfield  {journal} {\bibinfo
  {journal} {Phys. Rev. B}\ }\textbf {\bibinfo {volume} {76}},\ \bibinfo
  {pages} {245319} (\bibinfo {year} {2007})}\BibitemShut {NoStop}%
\bibitem [{\citenamefont {P\"oltl}\ \emph {et~al.}(2009)\citenamefont
  {P\"oltl}, \citenamefont {Emary},\ and\ \citenamefont {Brandes}}]{Poltl2009}%
  \BibitemOpen
  \bibfield  {author} {\bibinfo {author} {\bibfnamefont {C.}~\bibnamefont
  {P\"oltl}}, \bibinfo {author} {\bibfnamefont {C.}~\bibnamefont {Emary}},\
  and\ \bibinfo {author} {\bibfnamefont {T.}~\bibnamefont {Brandes}},\
  }\bibfield  {title} {\bibinfo {title} {Two-particle dark state in the
  transport through a triple quantum dot},\ }\href
  {https://doi.org/10.1103/PhysRevB.80.115313} {\bibfield  {journal} {\bibinfo
  {journal} {Phys. Rev. B}\ }\textbf {\bibinfo {volume} {80}},\ \bibinfo
  {pages} {115313} (\bibinfo {year} {2009})}\BibitemShut {NoStop}%
\bibitem [{\citenamefont {Kostyrko}\ and\ \citenamefont
  {Bu\l{}ka}(2009)}]{Kostyrko2009}%
  \BibitemOpen
  \bibfield  {author} {\bibinfo {author} {\bibfnamefont {T.}~\bibnamefont
  {Kostyrko}}\ and\ \bibinfo {author} {\bibfnamefont {B.~R.}\ \bibnamefont
  {Bu\l{}ka}},\ }\bibfield  {title} {\bibinfo {title} {Symmetry-controlled
  negative differential resistance effect in a triangular molecule},\ }\href
  {https://doi.org/10.1103/PhysRevB.79.075310} {\bibfield  {journal} {\bibinfo
  {journal} {Phys. Rev. B}\ }\textbf {\bibinfo {volume} {79}},\ \bibinfo
  {pages} {075310} (\bibinfo {year} {2009})}\BibitemShut {NoStop}%
\bibitem [{\citenamefont {Weymann}\ \emph {et~al.}(2011)\citenamefont
  {Weymann}, \citenamefont {Bu\l{}ka},\ and\ \citenamefont
  {Barna\ifmmode~\acute{s}\else \'{s}\fi{}}}]{Weymann2011}%
  \BibitemOpen
  \bibfield  {author} {\bibinfo {author} {\bibfnamefont {I.}~\bibnamefont
  {Weymann}}, \bibinfo {author} {\bibfnamefont {B.~R.}\ \bibnamefont
  {Bu\l{}ka}},\ and\ \bibinfo {author} {\bibfnamefont {J.}~\bibnamefont
  {Barna\ifmmode~\acute{s}\else \'{s}\fi{}}},\ }\bibfield  {title} {\bibinfo
  {title} {Dark states in transport through triple quantum dots: The role of
  cotunneling},\ }\href {https://doi.org/10.1103/PhysRevB.83.195302} {\bibfield
   {journal} {\bibinfo  {journal} {Phys. Rev. B}\ }\textbf {\bibinfo {volume}
  {83}},\ \bibinfo {pages} {195302} (\bibinfo {year} {2011})}\BibitemShut
  {NoStop}%
\bibitem [{\citenamefont {Wrzesniewski}\ and\ \citenamefont
  {Weymann}(2018)}]{Wrzesniewski2018}%
  \BibitemOpen
  \bibfield  {author} {\bibinfo {author} {\bibfnamefont {K.}~\bibnamefont
  {Wrzesniewski}}\ and\ \bibinfo {author} {\bibfnamefont {I.}~\bibnamefont
  {Weymann}},\ }\bibfield  {title} {\bibinfo {title} {Dark states in
  spin-polarized transport through triple quantum dot molecules},\ }\href
  {https://doi.org/10.1103/PhysRevB.97.075425} {\bibfield  {journal} {\bibinfo
  {journal} {Physical Review B}\ }\textbf {\bibinfo {volume} {97}},\ \bibinfo
  {pages} {075425} (\bibinfo {year} {2018})}\BibitemShut {NoStop}%
\bibitem [{\citenamefont {Pozner}\ \emph {et~al.}(2015)\citenamefont {Pozner},
  \citenamefont {Lifshitz},\ and\ \citenamefont {Peskin}}]{Pozner2015}%
  \BibitemOpen
  \bibfield  {author} {\bibinfo {author} {\bibfnamefont {R.}~\bibnamefont
  {Pozner}}, \bibinfo {author} {\bibfnamefont {E.}~\bibnamefont {Lifshitz}},\
  and\ \bibinfo {author} {\bibfnamefont {U.}~\bibnamefont {Peskin}},\
  }\bibfield  {title} {\bibinfo {title} {Negative differential resistance probe
  for interdot interactions in a double quantum dot array},\ }\href
  {https://doi.org/10.1021/acs.jpclett.5b00434} {\bibfield  {journal} {\bibinfo
   {journal} {The Journal of Physical Chemistry Letters}\ }\textbf {\bibinfo
  {volume} {6}},\ \bibinfo {pages} {1521} (\bibinfo {year} {2015})}\BibitemShut
  {NoStop}%
\bibitem [{\citenamefont {Ladr\'on~de Guevara}\ and\ \citenamefont
  {Orellana}(2006)}]{a01}%
  \BibitemOpen
  \bibfield  {author} {\bibinfo {author} {\bibfnamefont {M.~L.}\ \bibnamefont
  {Ladr\'on~de Guevara}}\ and\ \bibinfo {author} {\bibfnamefont {P.~A.}\
  \bibnamefont {Orellana}},\ }\bibfield  {title} {\bibinfo {title} {Electronic
  transport through a parallel-coupled triple quantum dot molecule: Fano
  resonances and bound states in the continuum},\ }\href
  {https://doi.org/10.1103/PhysRevB.73.205303} {\bibfield  {journal} {\bibinfo
  {journal} {Phys. Rev. B}\ }\textbf {\bibinfo {volume} {73}},\ \bibinfo
  {pages} {205303} (\bibinfo {year} {2006})}\BibitemShut {NoStop}%
\bibitem [{\citenamefont {Kwapi{\'{n}}ski}\ and\ \citenamefont
  {Taranko}(2014)}]{a02}%
  \BibitemOpen
  \bibfield  {author} {\bibinfo {author} {\bibfnamefont {T.}~\bibnamefont
  {Kwapi{\'{n}}ski}}\ and\ \bibinfo {author} {\bibfnamefont {R.}~\bibnamefont
  {Taranko}},\ }\bibfield  {title} {\bibinfo {title} {Charging time effects and
  transient current beats in horizontal and vertical quantum dot systems},\
  }\href {https://doi.org/10.1016/j.physe.2014.06.006} {\bibfield  {journal}
  {\bibinfo  {journal} {Physica E: Low-dimensional Systems and Nanostructures}\
  }\textbf {\bibinfo {volume} {63}},\ \bibinfo {pages} {241–247} (\bibinfo
  {year} {2014})}\BibitemShut {NoStop}%
\bibitem [{\citenamefont {Kohler}\ \emph {et~al.}(2005)\citenamefont {Kohler},
  \citenamefont {Lehmann},\ and\ \citenamefont {Hänggi}}]{a07}%
  \BibitemOpen
  \bibfield  {author} {\bibinfo {author} {\bibfnamefont {S.}~\bibnamefont
  {Kohler}}, \bibinfo {author} {\bibfnamefont {J.}~\bibnamefont {Lehmann}},\
  and\ \bibinfo {author} {\bibfnamefont {P.}~\bibnamefont {Hänggi}},\
  }\bibfield  {title} {\bibinfo {title} {Driven quantum transport on the
  nanoscale},\ }\href
  {https://doi.org/https://doi.org/10.1016/j.physrep.2004.11.002} {\bibfield
  {journal} {\bibinfo  {journal} {Physics Reports}\ }\textbf {\bibinfo {volume}
  {406}},\ \bibinfo {pages} {379 } (\bibinfo {year} {2005})}\BibitemShut
  {NoStop}%
\bibitem [{\citenamefont {Brandes}(2005)}]{Brandes2005}%
  \BibitemOpen
  \bibfield  {author} {\bibinfo {author} {\bibfnamefont {T.}~\bibnamefont
  {Brandes}},\ }\bibfield  {title} {\bibinfo {title} {Coherent and collective
  quantum optical effects in mesoscopic systems},\ }\href
  {https://doi.org/https://doi.org/10.1016/j.physrep.2004.12.002} {\bibfield
  {journal} {\bibinfo  {journal} {Physics Reports}\ }\textbf {\bibinfo {volume}
  {408}},\ \bibinfo {pages} {315} (\bibinfo {year} {2005})}\BibitemShut
  {NoStop}%
\bibitem [{\citenamefont {Volk}\ \emph {et~al.}(2013)\citenamefont {Volk},
  \citenamefont {Neumann}, \citenamefont {Kazarski}, \citenamefont {Fringes},
  \citenamefont {Engels}, \citenamefont {Haupt}, \citenamefont {Müller},\ and\
  \citenamefont {Stampfer}}]{Volk2013}%
  \BibitemOpen
  \bibfield  {author} {\bibinfo {author} {\bibfnamefont {C.}~\bibnamefont
  {Volk}}, \bibinfo {author} {\bibfnamefont {C.}~\bibnamefont {Neumann}},
  \bibinfo {author} {\bibfnamefont {S.}~\bibnamefont {Kazarski}}, \bibinfo
  {author} {\bibfnamefont {S.}~\bibnamefont {Fringes}}, \bibinfo {author}
  {\bibfnamefont {S.}~\bibnamefont {Engels}}, \bibinfo {author} {\bibfnamefont
  {F.}~\bibnamefont {Haupt}}, \bibinfo {author} {\bibfnamefont
  {A.}~\bibnamefont {Müller}},\ and\ \bibinfo {author} {\bibfnamefont
  {C.}~\bibnamefont {Stampfer}},\ }\bibfield  {title} {\bibinfo {title}
  {Probing relaxation times in graphene quantum dots},\ }\href
  {https://doi.org/10.1038/ncomms2738} {\bibfield  {journal} {\bibinfo
  {journal} {Nature Communications}\ }\textbf {\bibinfo {volume} {4}},\
  \bibinfo {pages} {1753} (\bibinfo {year} {2013})}\BibitemShut {NoStop}%
\bibitem [{\citenamefont {Jałochowski}\ \emph {et~al.}(2024)\citenamefont
  {Jałochowski}, \citenamefont {Krawiec},\ and\ \citenamefont
  {Kwapiński}}]{Jal2024}%
  \BibitemOpen
  \bibfield  {author} {\bibinfo {author} {\bibfnamefont {M.}~\bibnamefont
  {Jałochowski}}, \bibinfo {author} {\bibfnamefont {M.}~\bibnamefont
  {Krawiec}},\ and\ \bibinfo {author} {\bibfnamefont {T.}~\bibnamefont
  {Kwapiński}},\ }\bibfield  {title} {\bibinfo {title} {Implementation of the
  su–schrieffer–heeger model in the self-assembly si–in atomic chains on
  the si(553)–au surface},\ }\href {https://doi.org/10.1021/acsnano.4c00225}
  {\bibfield  {journal} {\bibinfo  {journal} {ACS Nano}\ }\textbf {\bibinfo
  {volume} {18}},\ \bibinfo {pages} {12861} (\bibinfo {year}
  {2024})}\BibitemShut {NoStop}%
\bibitem [{\citenamefont {Ja\l{}ochowski}\ and\ \citenamefont
  {Zdyb}(1997)}]{JAL1997}%
  \BibitemOpen
  \bibfield  {author} {\bibinfo {author} {\bibfnamefont {M.}~\bibnamefont
  {Ja\l{}ochowski}, \bibfnamefont {M.~Strozak}}\ and\ \bibinfo {author}
  {\bibfnamefont {R.}~\bibnamefont {Zdyb}},\ }\bibfield  {title} {\bibinfo
  {title} {Gold-induced ordering on vicinal si(111)},\ }\href
  {https://doi.org/https://doi.org/10.1016/S0039-6028(97)80009-X} {\bibfield
  {journal} {\bibinfo  {journal} {Surface Science}\ }\textbf {\bibinfo {volume}
  {375}},\ \bibinfo {pages} {203} (\bibinfo {year} {1997})}\BibitemShut
  {NoStop}%
\bibitem [{\citenamefont {Crain}\ \emph {et~al.}(2004)\citenamefont {Crain},
  \citenamefont {McChesney}, \citenamefont {Zheng}, \citenamefont {Gallagher},
  \citenamefont {Snijders}, \citenamefont {Bissen}, \citenamefont {Gundelach},
  \citenamefont {Erwin},\ and\ \citenamefont {Himpsel}}]{Crain2004}%
  \BibitemOpen
  \bibfield  {author} {\bibinfo {author} {\bibfnamefont {J.~N.}\ \bibnamefont
  {Crain}}, \bibinfo {author} {\bibfnamefont {J.~L.}\ \bibnamefont
  {McChesney}}, \bibinfo {author} {\bibfnamefont {F.}~\bibnamefont {Zheng}},
  \bibinfo {author} {\bibfnamefont {M.~C.}\ \bibnamefont {Gallagher}}, \bibinfo
  {author} {\bibfnamefont {P.~C.}\ \bibnamefont {Snijders}}, \bibinfo {author}
  {\bibfnamefont {M.}~\bibnamefont {Bissen}}, \bibinfo {author} {\bibfnamefont
  {C.}~\bibnamefont {Gundelach}}, \bibinfo {author} {\bibfnamefont {S.~C.}\
  \bibnamefont {Erwin}},\ and\ \bibinfo {author} {\bibfnamefont {F.~J.}\
  \bibnamefont {Himpsel}},\ }\bibfield  {title} {\bibinfo {title} {Chains of
  gold atoms with tailored electronic states},\ }\href
  {https://doi.org/10.1103/PhysRevB.69.125401} {\bibfield  {journal} {\bibinfo
  {journal} {Phys. Rev. B}\ }\textbf {\bibinfo {volume} {69}},\ \bibinfo
  {pages} {125401} (\bibinfo {year} {2004})}\BibitemShut {NoStop}%
\bibitem [{\citenamefont {Ahn}\ \emph {et~al.}(2005)\citenamefont {Ahn},
  \citenamefont {Kang}, \citenamefont {Ryang},\ and\ \citenamefont
  {Yeom}}]{Ahn2005}%
  \BibitemOpen
  \bibfield  {author} {\bibinfo {author} {\bibfnamefont {J.~R.}\ \bibnamefont
  {Ahn}}, \bibinfo {author} {\bibfnamefont {P.~G.}\ \bibnamefont {Kang}},
  \bibinfo {author} {\bibfnamefont {K.~D.}\ \bibnamefont {Ryang}},\ and\
  \bibinfo {author} {\bibfnamefont {H.~W.}\ \bibnamefont {Yeom}},\ }\bibfield
  {title} {\bibinfo {title} {Coexistence of two different peierls distortions
  within an atomic scale wire: Si(553)-au},\ }\bibfield  {journal} {\bibinfo
  {journal} {Physical Review Letters}\ }\textbf {\bibinfo {volume} {95}},\
  \href {https://doi.org/10.1103/physrevlett.95.196402}
  {10.1103/physrevlett.95.196402} (\bibinfo {year} {2005})\BibitemShut
  {NoStop}%
\bibitem [{\citenamefont {Crain}\ and\ \citenamefont
  {Pierce}(2005)}]{Crain2005}%
  \BibitemOpen
  \bibfield  {author} {\bibinfo {author} {\bibfnamefont {J.~N.}\ \bibnamefont
  {Crain}}\ and\ \bibinfo {author} {\bibfnamefont {D.~T.}\ \bibnamefont
  {Pierce}},\ }\bibfield  {title} {\bibinfo {title} {End states in
  one-dimensional atom chains},\ }\href
  {https://doi.org/10.1126/science.1106911} {\bibfield  {journal} {\bibinfo
  {journal} {Science}\ }\textbf {\bibinfo {volume} {307}},\ \bibinfo {pages}
  {703} (\bibinfo {year} {2005})},\ \Eprint
  {https://arxiv.org/abs/https://www.science.org/doi/pdf/10.1126/science.1106911}
  {https://www.science.org/doi/pdf/10.1126/science.1106911} \BibitemShut
  {NoStop}%
\bibitem [{\citenamefont {Snijders}\ \emph {et~al.}(2006)\citenamefont
  {Snijders}, \citenamefont {Rogge},\ and\ \citenamefont
  {Weitering}}]{Snijders2006}%
  \BibitemOpen
  \bibfield  {author} {\bibinfo {author} {\bibfnamefont {P.~C.}\ \bibnamefont
  {Snijders}}, \bibinfo {author} {\bibfnamefont {S.}~\bibnamefont {Rogge}},\
  and\ \bibinfo {author} {\bibfnamefont {H.~H.}\ \bibnamefont {Weitering}},\
  }\bibfield  {title} {\bibinfo {title} {Competing periodicities in
  fractionally filled one-dimensional bands},\ }\bibfield  {journal} {\bibinfo
  {journal} {Physical Review Letters}\ }\textbf {\bibinfo {volume} {96}},\
  \href {https://doi.org/10.1103/physrevlett.96.076801}
  {10.1103/physrevlett.96.076801} (\bibinfo {year} {2006})\BibitemShut
  {NoStop}%
\bibitem [{\citenamefont {Ryang}\ \emph {et~al.}(2007)\citenamefont {Ryang},
  \citenamefont {Kang}, \citenamefont {Yeom},\ and\ \citenamefont
  {Jeong}}]{Ryang2007}%
  \BibitemOpen
  \bibfield  {author} {\bibinfo {author} {\bibfnamefont {K.-D.}\ \bibnamefont
  {Ryang}}, \bibinfo {author} {\bibfnamefont {P.~G.}\ \bibnamefont {Kang}},
  \bibinfo {author} {\bibfnamefont {H.~W.}\ \bibnamefont {Yeom}},\ and\
  \bibinfo {author} {\bibfnamefont {S.}~\bibnamefont {Jeong}},\ }\bibfield
  {title} {\bibinfo {title} {Structures and defects of atomic wires on
  si(553)-au: An {STM} and theoretical study},\ }\bibfield  {journal} {\bibinfo
   {journal} {Physical Review B}\ }\textbf {\bibinfo {volume} {76}},\ \href
  {https://doi.org/10.1103/physrevb.76.205325} {10.1103/physrevb.76.205325}
  (\bibinfo {year} {2007})\BibitemShut {NoStop}%
\bibitem [{\citenamefont {Polei}\ \emph {et~al.}(2013)\citenamefont {Polei},
  \citenamefont {Snijders}, \citenamefont {Erwin}, \citenamefont {Himpsel},
  \citenamefont {Meiwes-Broer},\ and\ \citenamefont {Barke}}]{Polei2013}%
  \BibitemOpen
  \bibfield  {author} {\bibinfo {author} {\bibfnamefont {S.}~\bibnamefont
  {Polei}}, \bibinfo {author} {\bibfnamefont {P.~C.}\ \bibnamefont {Snijders}},
  \bibinfo {author} {\bibfnamefont {S.~C.}\ \bibnamefont {Erwin}}, \bibinfo
  {author} {\bibfnamefont {F.~J.}\ \bibnamefont {Himpsel}}, \bibinfo {author}
  {\bibfnamefont {K.-H.}\ \bibnamefont {Meiwes-Broer}},\ and\ \bibinfo {author}
  {\bibfnamefont {I.}~\bibnamefont {Barke}},\ }\bibfield  {title} {\bibinfo
  {title} {Structural transition in atomic chains driven by transient doping},\
  }\href {https://doi.org/10.1103/PhysRevLett.111.156801} {\bibfield  {journal}
  {\bibinfo  {journal} {Physical Review Letters}\ }\textbf {\bibinfo {volume}
  {111}},\ \bibinfo {pages} {156801} (\bibinfo {year} {2013})}\BibitemShut
  {NoStop}%
\bibitem [{\citenamefont {Hafke}\ \emph {et~al.}(2016)\citenamefont {Hafke},
  \citenamefont {Frigge}, \citenamefont {Witte}, \citenamefont {Krenzer},
  \citenamefont {Aulbach}, \citenamefont {Sch\"{a}fer}, \citenamefont
  {Claessen}, \citenamefont {Erwin},\ and\ \citenamefont {von
  Hoegen}}]{Hafke2016}%
  \BibitemOpen
  \bibfield  {author} {\bibinfo {author} {\bibfnamefont {B.}~\bibnamefont
  {Hafke}}, \bibinfo {author} {\bibfnamefont {T.}~\bibnamefont {Frigge}},
  \bibinfo {author} {\bibfnamefont {T.}~\bibnamefont {Witte}}, \bibinfo
  {author} {\bibfnamefont {B.}~\bibnamefont {Krenzer}}, \bibinfo {author}
  {\bibfnamefont {J.}~\bibnamefont {Aulbach}}, \bibinfo {author} {\bibfnamefont
  {J.}~\bibnamefont {Sch\"{a}fer}}, \bibinfo {author} {\bibfnamefont
  {R.}~\bibnamefont {Claessen}}, \bibinfo {author} {\bibfnamefont {S.~C.}\
  \bibnamefont {Erwin}},\ and\ \bibinfo {author} {\bibfnamefont {M.~H.}\
  \bibnamefont {von Hoegen}},\ }\bibfield  {title} {\bibinfo {title}
  {Two-dimensional interaction of spin chains in the si(553)-au nanowire
  system},\ }\bibfield  {journal} {\bibinfo  {journal} {Physical Review B}\
  }\textbf {\bibinfo {volume} {94}},\ \href
  {https://doi.org/10.1103/physrevb.94.161403} {10.1103/physrevb.94.161403}
  (\bibinfo {year} {2016})\BibitemShut {NoStop}%
\bibitem [{\citenamefont {Braun}\ \emph {et~al.}(2018)\citenamefont {Braun},
  \citenamefont {Gerstmann},\ and\ \citenamefont {Schmidt}}]{Braun2018}%
  \BibitemOpen
  \bibfield  {author} {\bibinfo {author} {\bibfnamefont {C.}~\bibnamefont
  {Braun}}, \bibinfo {author} {\bibfnamefont {U.}~\bibnamefont {Gerstmann}},\
  and\ \bibinfo {author} {\bibfnamefont {W.~G.}\ \bibnamefont {Schmidt}},\
  }\bibfield  {title} {\bibinfo {title} {Spin pairing versus spin chains at
  si(553)-au surfaces},\ }\bibfield  {journal} {\bibinfo  {journal} {Physical
  Review B}\ }\textbf {\bibinfo {volume} {98}},\ \href
  {https://doi.org/10.1103/physrevb.98.121402} {10.1103/physrevb.98.121402}
  (\bibinfo {year} {2018})\BibitemShut {NoStop}%
\bibitem [{\citenamefont {Edler}\ \emph {et~al.}(2019)\citenamefont {Edler},
  \citenamefont {Miccoli}, \citenamefont {Pfn\"ur},\ and\ \citenamefont
  {Tegenkamp}}]{Edler2019}%
  \BibitemOpen
  \bibfield  {author} {\bibinfo {author} {\bibfnamefont {F.}~\bibnamefont
  {Edler}}, \bibinfo {author} {\bibfnamefont {I.}~\bibnamefont {Miccoli}},
  \bibinfo {author} {\bibfnamefont {H.}~\bibnamefont {Pfn\"ur}},\ and\ \bibinfo
  {author} {\bibfnamefont {C.}~\bibnamefont {Tegenkamp}},\ }\bibfield  {title}
  {\bibinfo {title} {Charge-transfer transition in au-induced quantum wires on
  si(553)},\ }\href {https://doi.org/10.1103/PhysRevB.100.045419} {\bibfield
  {journal} {\bibinfo  {journal} {Phys. Rev. B}\ }\textbf {\bibinfo {volume}
  {100}},\ \bibinfo {pages} {045419} (\bibinfo {year} {2019})}\BibitemShut
  {NoStop}%
\bibitem [{\citenamefont {Mamiyev}\ \emph {et~al.}(2021)\citenamefont
  {Mamiyev}, \citenamefont {Fink}, \citenamefont {Holtgrewe}, \citenamefont
  {Pfn\"ur},\ and\ \citenamefont {Sanna}}]{Pfnur2021}%
  \BibitemOpen
  \bibfield  {author} {\bibinfo {author} {\bibfnamefont {Z.}~\bibnamefont
  {Mamiyev}}, \bibinfo {author} {\bibfnamefont {C.}~\bibnamefont {Fink}},
  \bibinfo {author} {\bibfnamefont {K.}~\bibnamefont {Holtgrewe}}, \bibinfo
  {author} {\bibfnamefont {H.}~\bibnamefont {Pfn\"ur}},\ and\ \bibinfo {author}
  {\bibfnamefont {S.}~\bibnamefont {Sanna}},\ }\bibfield  {title} {\bibinfo
  {title} {Enforced long-range order in 1d wires by coupling to higher
  dimensions},\ }\href {https://doi.org/10.1103/PhysRevLett.126.106101}
  {\bibfield  {journal} {\bibinfo  {journal} {Phys. Rev. Lett.}\ }\textbf
  {\bibinfo {volume} {126}},\ \bibinfo {pages} {106101} (\bibinfo {year}
  {2021})}\BibitemShut {NoStop}%
\bibitem [{\citenamefont {Krawiec}(2010)}]{Krawiec2010}%
  \BibitemOpen
  \bibfield  {author} {\bibinfo {author} {\bibfnamefont {M.}~\bibnamefont
  {Krawiec}},\ }\bibfield  {title} {\bibinfo {title} {Structural model of the
  au-induced si(553) surface: Double au rows},\ }\bibfield  {journal} {\bibinfo
   {journal} {Physical Review B}\ }\textbf {\bibinfo {volume} {81}},\ \href
  {https://doi.org/10.1103/physrevb.81.115436} {10.1103/physrevb.81.115436}
  (\bibinfo {year} {2010})\BibitemShut {NoStop}%
\bibitem [{Yeo(2022)}]{Yeom2022}%
  \BibitemOpen
  \ \href {https://doi.org/doi.org/10.1021/acsnano.2c00972}
  {doi.org/10.1021/acsnano.2c00972} (\bibinfo {year} {2022})\BibitemShut
  {NoStop}%
\bibitem [{\citenamefont {Jałochowski}\ and\ \citenamefont
  {Kwapiński}(2023)}]{Jal2023}%
  \BibitemOpen
  \bibfield  {author} {\bibinfo {author} {\bibfnamefont {M.}~\bibnamefont
  {Jałochowski}}\ and\ \bibinfo {author} {\bibfnamefont {T.}~\bibnamefont
  {Kwapiński}},\ }\bibfield  {title} {\bibinfo {title} {Distribution of
  electron density in self-assembled one-dimensional chains of si atoms},\
  }\bibfield  {journal} {\bibinfo  {journal} {Materials}\ }\textbf {\bibinfo
  {volume} {16}},\ \href {https://doi.org/10.3390/ma16176044}
  {10.3390/ma16176044} (\bibinfo {year} {2023})\BibitemShut {NoStop}%
\bibitem [{\citenamefont {Park}\ \emph {et~al.}(2022)\citenamefont {Park},
  \citenamefont {Do}, \citenamefont {Shin}, \citenamefont {Song}, \citenamefont
  {Stetsovych}, \citenamefont {Jelinek},\ and\ \citenamefont
  {Yeom}}]{Yeom2022NN}%
  \BibitemOpen
  \bibfield  {author} {\bibinfo {author} {\bibfnamefont {J.~W.}\ \bibnamefont
  {Park}}, \bibinfo {author} {\bibfnamefont {E.}~\bibnamefont {Do}}, \bibinfo
  {author} {\bibfnamefont {J.~S.}\ \bibnamefont {Shin}}, \bibinfo {author}
  {\bibfnamefont {S.~K.}\ \bibnamefont {Song}}, \bibinfo {author}
  {\bibfnamefont {O.}~\bibnamefont {Stetsovych}}, \bibinfo {author}
  {\bibfnamefont {P.}~\bibnamefont {Jelinek}},\ and\ \bibinfo {author}
  {\bibfnamefont {H.~W.}\ \bibnamefont {Yeom}},\ }\bibfield  {title} {\bibinfo
  {title} {Creation and annihilation of mobile fractional solitons in atomic
  chains},\ }\href {https://doi.org/10.1038/s41565-021-01042-8} {\bibfield
  {journal} {\bibinfo  {journal} {Nature Nanotechnology}\ }\textbf {\bibinfo
  {volume} {17}},\ \bibinfo {pages} {244} (\bibinfo {year} {2022})}\BibitemShut
  {NoStop}%
\bibitem [{\citenamefont {Snijders1}\ \emph {et~al.}(2012)\citenamefont
  {Snijders1}, \citenamefont {Johnson}, \citenamefont {Guisinger},
  \citenamefont {Erwin},\ and\ \citenamefont {Himpsel}}]{Snijders2012}%
  \BibitemOpen
  \bibfield  {author} {\bibinfo {author} {\bibfnamefont {P.~C.}\ \bibnamefont
  {Snijders1}}, \bibinfo {author} {\bibfnamefont {P.~S.}\ \bibnamefont
  {Johnson}}, \bibinfo {author} {\bibfnamefont {N.~P.}\ \bibnamefont
  {Guisinger}}, \bibinfo {author} {\bibfnamefont {S.~C.}\ \bibnamefont
  {Erwin}},\ and\ \bibinfo {author} {\bibfnamefont {F.~J.}\ \bibnamefont
  {Himpsel}},\ }\bibfield  {title} {\bibinfo {title} {Spectroscopic evidence
  for spin-polarized edge states in graphitic si nanowires},\ }\href
  {https://doi.org/10.1088/1367-2630/14/10/103004} {\bibfield  {journal}
  {\bibinfo  {journal} {New Journal of Physics}\ }\textbf {\bibinfo {volume}
  {14}},\ \bibinfo {pages} {103004} (\bibinfo {year} {2012})}\BibitemShut
  {NoStop}%
\bibitem [{\citenamefont {Datta}(1995)}]{a09}%
  \BibitemOpen
  \bibfield  {author} {\bibinfo {author} {\bibfnamefont {S.}~\bibnamefont
  {Datta}},\ }\href {https://doi.org/10.1017/CBO9780511805776} {\emph {\bibinfo
  {title} {Electronic Transport in Mesoscopic Systems}}},\ Cambridge Studies in
  Semiconductor Physics and Microelectronic Engineering\ (\bibinfo  {publisher}
  {Cambridge University Press},\ \bibinfo {year} {1995})\BibitemShut {NoStop}%
\bibitem [{\citenamefont {Kurzyna}\ and\ \citenamefont
  {Kwapiński}(2018)}]{a10}%
  \BibitemOpen
  \bibfield  {author} {\bibinfo {author} {\bibfnamefont {M.}~\bibnamefont
  {Kurzyna}}\ and\ \bibinfo {author} {\bibfnamefont {T.}~\bibnamefont
  {Kwapiński}},\ }\bibfield  {title} {\bibinfo {title} {Non-local electron
  transport through normal and topological ladder-like atomic systems},\ }\href
  {https://doi.org/10.1063/1.5028571} {\bibfield  {journal} {\bibinfo
  {journal} {Journal of Applied Physics}\ }\textbf {\bibinfo {volume} {123}},\
  \bibinfo {pages} {194301} (\bibinfo {year} {2018})},\ \Eprint
  {https://arxiv.org/abs/https://doi.org/10.1063/1.5028571}
  {https://doi.org/10.1063/1.5028571} \BibitemShut {NoStop}%
\bibitem [{\citenamefont {Grimley}\ \emph {et~al.}(1983)\citenamefont
  {Grimley}, \citenamefont {Jyothi~Bhasu},\ and\ \citenamefont
  {Sebastian}}]{Grim}%
  \BibitemOpen
  \bibfield  {author} {\bibinfo {author} {\bibfnamefont {T.}~\bibnamefont
  {Grimley}}, \bibinfo {author} {\bibfnamefont {V.}~\bibnamefont
  {Jyothi~Bhasu}},\ and\ \bibinfo {author} {\bibfnamefont {K.}~\bibnamefont
  {Sebastian}},\ }\bibfield  {title} {\bibinfo {title} {Electron transfer in
  the reflection of atoms from metal surfaces},\ }\href
  {https://doi.org/10.1016/0039-6028(83)90352-7} {\bibfield  {journal}
  {\bibinfo  {journal} {Surf.Sci.}\ }\textbf {\bibinfo {volume} {121}},\
  \bibinfo {pages} {305} (\bibinfo {year} {1983})}\BibitemShut {NoStop}%
\bibitem [{\citenamefont {Kwapiński}\ and\ \citenamefont
  {Taranko}(2003)}]{Tar2003}%
  \BibitemOpen
  \bibfield  {author} {\bibinfo {author} {\bibfnamefont {T.}~\bibnamefont
  {Kwapiński}}\ and\ \bibinfo {author} {\bibfnamefont {R.}~\bibnamefont
  {Taranko}},\ }\bibfield  {title} {\bibinfo {title} {Time-dependent transport
  through a quantum dot with the over-dot (bridge) additional tunneling
  channel},\ }\href
  {https://doi.org/https://doi.org/10.1016/S1386-9477(02)01105-0} {\bibfield
  {journal} {\bibinfo  {journal} {Physica E: Low-dimensional Systems and
  Nanostructures}\ }\textbf {\bibinfo {volume} {18}},\ \bibinfo {pages} {402}
  (\bibinfo {year} {2003})}\BibitemShut {NoStop}%
\bibitem [{\citenamefont {Zhou}\ \emph {et~al.}(2008)\citenamefont {Zhou},
  \citenamefont {Wang}, \citenamefont {Sheng}, \citenamefont {Wang},\ and\
  \citenamefont {Xing}}]{Zhou2008}%
  \BibitemOpen
  \bibfield  {author} {\bibinfo {author} {\bibfnamefont {Y.-Q.}\ \bibnamefont
  {Zhou}}, \bibinfo {author} {\bibfnamefont {R.-Q.}\ \bibnamefont {Wang}},
  \bibinfo {author} {\bibfnamefont {L.}~\bibnamefont {Sheng}}, \bibinfo
  {author} {\bibfnamefont {B.}~\bibnamefont {Wang}},\ and\ \bibinfo {author}
  {\bibfnamefont {D.~Y.}\ \bibnamefont {Xing}},\ }\bibfield  {title} {\bibinfo
  {title} {Pumped spin and charge currents from applying a microwave field to a
  quantum dot between two magnetic leads},\ }\href
  {https://doi.org/10.1103/PhysRevB.78.155327} {\bibfield  {journal} {\bibinfo
  {journal} {Phys. Rev. B}\ }\textbf {\bibinfo {volume} {78}},\ \bibinfo
  {pages} {155327} (\bibinfo {year} {2008})}\BibitemShut {NoStop}%
\bibitem [{\citenamefont {Kurzyna}\ and\ \citenamefont
  {Kwapi\ifmmode~\acute{n}\else \'{n}\fi{}ski}(2020)}]{Kwap2020}%
  \BibitemOpen
  \bibfield  {author} {\bibinfo {author} {\bibfnamefont {M.}~\bibnamefont
  {Kurzyna}}\ and\ \bibinfo {author} {\bibfnamefont {T.}~\bibnamefont
  {Kwapi\ifmmode~\acute{n}\else \'{n}\fi{}ski}},\ }\bibfield  {title} {\bibinfo
  {title} {Nontrivial dynamics of a two-site system: Transient crystals},\
  }\href {https://doi.org/10.1103/PhysRevB.102.245414} {\bibfield  {journal}
  {\bibinfo  {journal} {Phys. Rev. B}\ }\textbf {\bibinfo {volume} {102}},\
  \bibinfo {pages} {245414} (\bibinfo {year} {2020})}\BibitemShut {NoStop}%
\bibitem [{\citenamefont {Kwapi\ifmmode~\acute{n}\else
  \'{n}\fi{}ski}(2023)}]{Kwap2023}%
  \BibitemOpen
  \bibfield  {author} {\bibinfo {author} {\bibfnamefont {T.}~\bibnamefont
  {Kwapi\ifmmode~\acute{n}\else \'{n}\fi{}ski}},\ }\bibfield  {title} {\bibinfo
  {title} {One-dimensional transient crystals},\ }\href
  {https://doi.org/10.1103/PhysRevB.107.035422} {\bibfield  {journal} {\bibinfo
   {journal} {Phys. Rev. B}\ }\textbf {\bibinfo {volume} {107}},\ \bibinfo
  {pages} {035422} (\bibinfo {year} {2023})}\BibitemShut {NoStop}%
\end{thebibliography}%

\end{document}